\documentclass[twocolumn,showpacs,preprintnumbers,amsmath,amssymb,prl]{revtex4}

\usepackage{graphicx}
\usepackage{bm}

\begin{document}

\preprint{}

\title{Improving immunization strategies}
\author{Lazaros K. Gallos$^1$}
\author{Fredrik Liljeros$^2$}
\author{Panos Argyrakis$^1$}
\author{Armin Bunde$^3$}
\author{Shlomo Havlin$^4$}
\affiliation{$^1$Department of Physics, University of Thessaloniki, 54124 Thessaloniki, Greece}
\affiliation{$^2$Department of Sociology, Stockholm University 106 91 Stockholm, Sweden}
\affiliation{$^3$Institut f\"{u}r Theoretische Physik III, Justus-Liebig-Universit\"{a}t Giessen, Heinrich-Buff-Ring 16, 35392 Giessen, Germany}
\affiliation{$^4$Minerva Center and Department of Physics, Bar-Ilan University, 52900 Ramat-Gan, Israel}

\date{\today}

\begin{abstract}
We introduce an immunization method where the percentage of required vaccinations for
immunity are close to the optimal value of a targeted immunization scheme of highest degree nodes. 
Our strategy retains the advantage of being purely local,
without the need of knowledge on the global network structure or identification of the highest degree nodes. The
method consists of selecting a random node and asking for a neighbor that has more links than himself
or more than a given threshold and immunizing him. We compare this
method to other efficient strategies on three real social networks and on a scale-free network model, and
find it to be significantly more effective.
\end{abstract}


\maketitle
Immunization of large populations through vaccination is an extremely important
issue with obvious implications for the public health \cite{Anderson, Britton, Ball}. The eradication of
Small Pox through a global mass vaccination campaign during the second part of
the 20th century represents, for example, a landmark in the history of the medical
sciences \cite{Bazin}. Global or national mass vaccination may however not always be
possible. The number of vaccinated people may need to be minimized due to
severe side effects of vaccination such as for Small Pox, or temporary shortage of vaccine that could be
the case for a pandemic influenza. The cost for a vaccine may also be an
important limiting factor. Improving efficiency of immunization is thus an
urgent task.

Recently \cite{AB02}, developments in the study of population connectivities helped researchers in the field to present
new ideas on immunization, based on the heterogeneity in the number of contacts between individuals.
A number of strategies have
been proposed for lowering the required minimum fraction $f_c$ of the population to be immunized.
The problem can be mapped to the well-known percolation problem where nodes are immunized (removed)
up to a concentration $f_c$, above which the spanning cluster does not survive. Random immunization of
nodes has been shown incapable of protecting the population when the contacts distribution is wide, since the
percolation threshold is close to $f_c=1$, i.e. practically all nodes need to be immunized
\cite{Cohen00,Callaway,Vespignani}.
The best known strategy today is believed to be targeted immunization, where the highest connected
nodes in the system are immunized in decreasing order of their degree. In this case $f_c$ is less than 10\%
\cite{Barabasi,Cohen01,Callaway}. For all practical applications, though, this
approach is unrealistic because it is a `global' strategy and requires a complete knowledge of the high
degree nodes,
which is in many cases impossible. An effective strategy, called acquaintance
immunization, was recently introduced \cite{Cohen03} that combines both efficiency and somewhat greater ease of applicability.
According to this scheme a random individual is selected who then points to one of his random acquaintances
and this node is the one to be immunized.
This method is more efficient compared to random immunization ($f_c$ is of the order of 20-25\%) but less efficient than
targeted immunization.

In this paper we introduce an immunization method, which is practically as efficient as
the accepted as optimum strategy, but at the same time depends on local information only.
The method consists in selecting
random individuals and asking them to direct us to their friend who is more connected than
they are and this acuaintance is immunized. If such a friend does not exist we continue with another random selection.
Alternatively, in a second variation of the method we ask the randomly chosen individual to point us to a random neighbor that has a number of 
neighbors larger than e.g. $k=5$ (or an equally small and easily countable threshold value). If they point to such an
individual it is immunized, otherwise we select another individual.
Similar results are obtained if the chosen individual is asked to estimate his
own number of contacts, rather than of his random neighbor. Although this procedure is simpler, the selection
of a neighbor can also eliminate the bias that may be introduced due to selfish people, lying about their contacts
in order to receive the vaccine themselves. The method is proposed for social networks, but
it is expected that it can be even more efficient for technological networks, such as e.g. the Internet,
where the number of links for a given node is exactly known to the local network administrator,
and need not be estimated.

Our method is local because the decision for immunization of a given node
is taken without the need to know the connectivity of other nodes. This is in
contrast with global strategies where immunization of a node has to be decided
only after we have gathered information for the entire network. This means that
for immunization of e.g. a city or a country in a global method we have to send
special teams to collect this information and transmit it to a central place. This
central authority decides then which nodes should be immunized and transmits
back the outcome to the local authorities which then go on with vaccinations.
For a local method, there is no need to collect or compare data from other
areas of the network. Based on the answer of each individual the decision is
made immediately on whether a node should be immunized or not.

We study the proposed method on real social networks with a fat tail in their degree distribution, as well as on a random scale-free model network.
We also compare this method with several other immunization strategies, including such that partial knowledge on the global
network of contacts is available and we demonstrate the advantage of the proposed method
via the improvement in $f_c$.

The social networks used in this study represent different interactions among the members of an online community,
as described in Ref.~\cite{Holme}. These interactions include a) exchange of messages, b) signing of guestbooks,
c) flirt requests, and d) established friendships. The first three networks are directed
but we consider only their undirected projection, by transforming arcs into edges.
No significant difference is observed in the results for the undirected network and the
projections of the directed networks.
The size of the networks is of the order $N=10^4$. The percentage of immunized nodes is denoted with $f$,
while the percentage of nodes suveyed is denoted with $p$.
The four strategies that we employ are summarized below.
{\it Strategy I}: Immunize a node with probability proportional to $k^\alpha$, where $k$ is the number of connections
and $\alpha$ tunes the probability of preferentially selecting high-connectivity
or low-connectivity nodes. Large positive values of $\alpha$ tend towards mainly selecting the hubs
($\alpha\to \infty$ is equivalent to targeted immunization),
the value $\alpha=0$ represents the random immunization model, while negative $\alpha$
values lead to selecting the lower-connected nodes \cite{Gallos}. This parameter can be interpreted as a measure of
the extent of our knowledge on the structure. {\it Strategy II}: Select a node with probability proportional to $k^\alpha$
and immunize a random acquaintance of this node. The value $\alpha=0$ corresponds to
the acquaintance immunization scheme \cite{Cohen03}. {\it Strategy III}: Select a random node and immunize one of
its acquaintances $i$, with probability proportional to $k_i^\alpha$, where $k_i$ represents the degree of the neighbor.
{\it Strategy IV}: Select a random node and ask for an acquaintance, which is immunized if a certain condition is met. We study two variations:
a) The selected node points randomly to a node which is more connected than himself.
If there are no such neighbors no node is immunized.
b) The selected node is asked to choose a random neighbor with degree larger than a threshold value $k_{\rm cut}$
then this acquaintance is immunized.
Equivalently, we can ask the node to estimate its own degree. If it is larger than a threshold value we
immunize the node, otherwise we ignore it. These two variations are similar when $k_{\rm cut}=\langle k \rangle$.
We call strategy IV ``enhanced acquaintance immunization'' (EAI) method.

\begin{figure}
\includegraphics{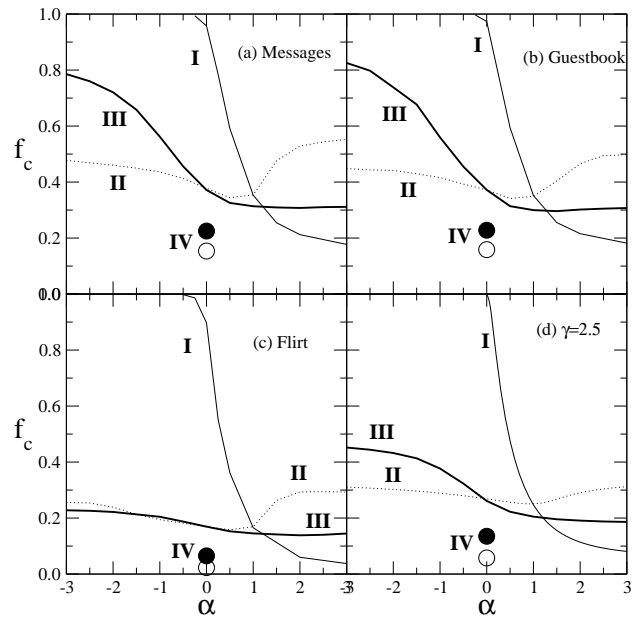}
\caption{\label{fig1} 
Critical immunized fraction $f_c$ of the population as a function of $\alpha$ for (a)-(c) 
Real-life social networks, and (d) scale-free network model with $\gamma=2.5$. Four different
strategies are used as described in the text and indicated in the plot. The two symbols correspond to the critical fraction
for the strategies of the enhanced acquaintance immunization method (the open circle corresponds to asking for an acquintance with threshold $k_{\rm cut}=7$,
while the filled circle corresponds to asking for a better connected node).
}
\end{figure}


In Figs.~\ref{fig1}a-c we present the results of $f_c$ for the four described strategies applied to three of the social
networks, as defined by different types of interactions. All networks follow similar patterns for a given strategy. In strategy I
we can see the abrupt decrease of $f_c$ when increasing $\alpha$ from $\alpha\leq 0$ (random immunization) with $f_c=1$ to $\alpha=\infty$ (targeted
immunization) with $f_c\ll 1$. Strategy II presents an improvement over the first strategy for values $\alpha\lesssim 1$. The critical value
$f_c$ presents a minimum at $\alpha\simeq 1$, indicating that identification of large hubs actually deteriorates the results,
since the neighbors of large hubs, which are chosen to be immunized, are with higher probability low degree nodes for
dissasortative networks, similarly with the acquaintance immunization method \cite{Cohen03}.
Strategy III leads to monotonic decrease in $f_c$ and prevails from the first two methods when we have limited global network
knowledge, i.e. in the range $\alpha \in [0,1]$. However, in Strategy III we find that  when $\alpha=\infty$
(i.e. we always immunize the most connected neighbor) it may
be impossible to destroy the spanning cluster, because almost all selected nodes point to
the same hubs.
Finally, the enhanced acquaintance immunization strategy seems to be the most efficient method, although  it assumes no knowledge of the underlying structure
(the method is independent of $\alpha$). The value of $f_c$ is lower than
an attack with $\alpha=3$ and very close to the results of the targeted immunization.

To gain more insight into the different immunization methods we also performed numerical simulations on a model network.
We consider each member of a population represented by a node, while the acquaintances
of a person with other people form links. It is well established that many social networks follow a broad
distribution in the degree of a node, such as the power-law distribution $P(k) \sim k^{-\gamma}$,
where the exponent $\gamma$ is usually found to be between $2<\gamma <4$ \cite{AB02, Mendes, VespBook, Newman}. The above real networks are
scale-free with $\gamma\simeq 2.4$\cite{Holme}.
The results in Fig.~\ref{fig1}d correspond to the four strategies in such a model network (created with the configuration random model \cite{MR95})
with exponent $\gamma=2.5$, which is close
to the reported exponent $\gamma\simeq 2.4$ of the real networks used. All strategies in this plot follow
closely the results for the real networks.

\begin{figure}
\includegraphics{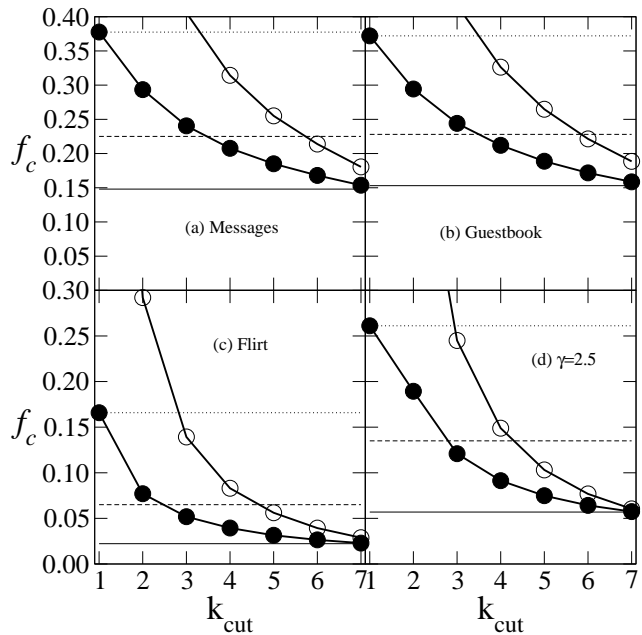}
\caption{\label{fig2}
Critical immunized fraction $f_c$ of the population as a function of the threshold value $k_{\rm cut}$ for
the enhanced acquaintance immunization strategy applied to (a)-(c) social interaction networks, (d) random model scale-free network with $\gamma=2.5$ (of
size $N=10^5$ nodes).
Filled symbols correspond to immunizing a
random neighbor of the selected node if its degree is $\geq k_{\rm cut}$ and open symbols to immunizing
the selected node itself. The upper horizontal dotted line is the result for acquaintance
immunization, the dashed line in the middle corresponds to immunizing a more connected acquaintance, while the lower line refers to targeted immunization.
}
\end{figure}

The two `transition' points for the first three strategies are located at $\alpha=0$ and $\alpha=1$.
At $\alpha=0$, strategies II and III coincide. In the range $\alpha\in [0,1]$ strategy III is more efficient,
indicating that in this range it is preferable to let the nodes choose their neighbors according to their
connectivity, rather than selecting nodes with probability proportional to $k^\alpha$ and following random links.
The value $\alpha\simeq 1$ is the optimum value for strategy II. In practice, the process is
equivalent to selecting a random link and immunizing one of the two nodes attached to the given link (provided
the uncorrelated network hypothesis holds). It is also interesting to note that up to the value $\alpha=1$ the
acquaintance immunization strategy is superior to direct immunization of the initially selected nodes, but close to this
value the two methods yield a similar value for $f_c$. When $\alpha>1$ the direct immunization method becomes more
efficient than acquaintance immunization.

\begin{figure}
\includegraphics{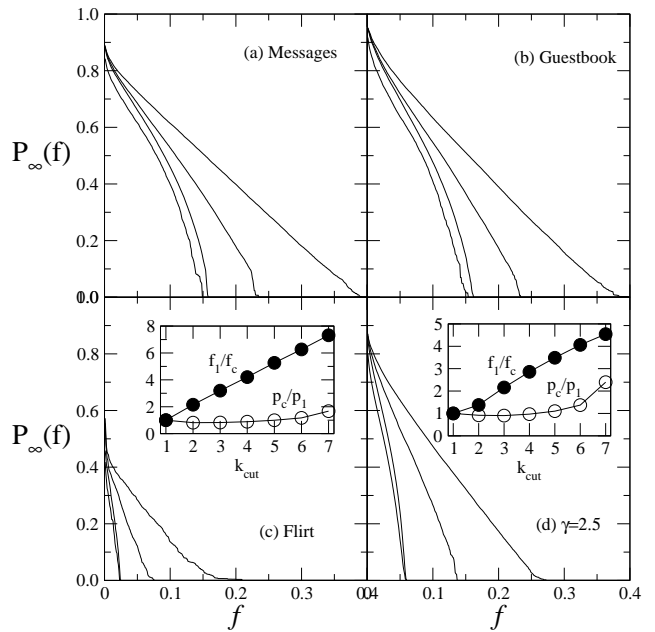}
\caption{\label{fig3} 
Size of epidemics, measured via the fraction of nodes belonging to the largest cluster over the number of not-immunized
nodes $P_\infty(f)$, as a function of the fraction $f$ of immunized nodes.
In each plot, from top to bottom, the curves correspond to acquaintance immunization, EAI redirecting to
a better connected node, the EAI with $k_{\rm cut}=7$, and
targeted immunization.
(a)-(c): real networks, and (d): random scale-free network with $\gamma=2.5$.
Insets: Ratios for $f_1/f_c$ of the critical immnunized fraction $f_c$ over the critical fraction $f_1$ for acquaintance
immunization ($k_{\rm cut}=1$) and $p_c/p_1$, i.e. the number of people surveyed, as a function of $k_{\rm cut}$ for
the EAI method.
}
\end{figure}

The enhanced acquaintance immunization is, however, found to be superior to all the above methods. The value of $f_c$
for a given $k_{\rm cut}$ value is of course independent of $\alpha$, meaning that it works equally well when there is no
further information on the network structure, i.e. global knowledge does not offer any significant advantage
over completely random selections. Thus, the strategy is local and easy to implement.
The choice of $k_{\rm cut}$, though, influences $f_c$ and can further reduce the $f_c$ value when more accurate
knowledge on the network structure is available.

The gain of this method for $k_{\rm cut}=7$ when compared to the
original acquaintance immunization method is about a factor of 4, which is for practical purposes a significant improvement.
This striking variation is evident in Fig.~\ref{fig2}, where the critical percentage decays from $f_c\simeq 0.26$ at
$k_{\rm cut}=1$ (acquaintance immunization) to $f_c\simeq 0.06$ at $k_{\rm cut}=7$. For $k_{\rm cut}=7$ the strategy works comparably well to the targeted
immunization. The fraction $f_c$, however, remains very low even when the cutoff value $k_{\rm cut}$ decreases to values close to,
but less than 7. This stability over the value of $k_{\rm cut}$ offers greater flexibility since the method seems tolerant to
mistakes of lower degree nodes being pointed at for immunization, without siginificant loss in the efficiency (even at a value
of $k_{\rm cut}=4$ the critical fraction $f_c$ remains lower than 10\%).
The results are different when we immunize directly the initially selected random node (without asking for an acquaintance)
and only at $k_{\rm cut}=7$ the two methods seem to coincide (Fig.~\ref{fig2}). There exists, though,  a critical degree
above which this strategy no longer works, simply because the number of nodes with degree larger than this value
is smaller than the critical number needed for complete immunization. Thus, it seems preferable to remain conservative
on the estimation of $k_{\rm cut}$ and choose a smaller value over a larger one.

A considerable advantage is gained, even when the question is posed in a much simpler way, i.e. we ask a random node to direct us to
a friend who is better connected than his and immunize him. This simple approach already offers a significant improvement over the original acquaintance method,
as is evident in Fig.~\ref{fig2}, although it is not as efficient as when asking for a friend whose degree exceeds the cutoff value.
Since it is, however, much easier for an individual to estimate an acquaintance who is better connected than himself, and practically
everyone can understand and correctly answer this simple question, we consider this method as a useful strategy which is easy to
apply in real-life situations.

In order to assess the size of the epidemics in the immunization process we measure the
size of the spanning cluster (epidemics size) as a function of the immunized
nodes $f$. In Figs.~\ref{fig3}a-c we present the fraction of nodes
belonging to the spanning cluster over the total number of non-immunized nodes for the real networks
described above and compare the targeted immunization with the enhanced acquaintance immunization and the original acquaintance immunization methods. The results for the model
scale-free networks (Fig.~\ref{fig3}d) are averages over 100 different realizations of networks with exponent $\gamma=2.5$.
In all cases the critical fraction for the targeted immunization and the EAI with the cutoff value are similar, while acquaintance immunization leads to considerably higher
values of $f_c$. Again, the EAI with an estimation of a better connected friend yields a result between these two extremes. However, during the removal process the targeted
immunization yields the faster decomposition of the spanning cluster, since it first removes the most connected nodes in the system.
The results for all the acquaintance immunization methods depend on when these largest hubs will be selected and the averaging conceals the fact that
during one realization the size of the largest cluster drops abruptly when the largest hubs are selected. Despite this, the proposed
methods follow closely the results of targeted immunization, while retaining the advantage of being local.

In the insets of Fig.~\ref{fig3} we can see that compared to the acquaintance immunization method (which is the EAI method with
$k_{\rm cut}=1$) in general we need to survey more nodes for their acquaintances
as $k_{\rm cut}$ increases, but this is a small change compared to the improvement in the number of required
immunizations presented in the same plots.

A work with similar scope was performed by Holme \cite{Holmeb}.
Among other methods, an immunization scheme was introduced, where a random
node points to one of its highest degree neighbors or to its most
connected neighbor. This corresponds to strategy III of the current work
with $\alpha\rightarrow\infty$ (where we encounter the problem of selecting always the same nodes
as described above) and the first variation of Strategy IV. The results
in that paper are consistent with the
ones presented above for these limiting cases.

In summary, we introduced and compared various immunization strategies on real and model networks. We have shown that the fraction
of immunized nodes can be significantly reduced to the almost optimum level
of intentional immunization using a completely local information strategy.
This simple
process is enough to ensure that the immunization threshold is significantly lowered, as compared to other local methods.

\begin{acknowledgments}
We thank R.~Cohen and D.~Brewer for useful discussions.
This work was supported by a European research NEST/PATHFINDER project DYSONET 012911, by a project of the Greek
GGET in conjunction with ESF and by the Israel Science Foundation.
\end{acknowledgments}

\end{document}